\documentclass[prb,twocolumn,aps,showpacs,floatfix,superscriptaddress]{revtex4}

\usepackage{graphicx}
\usepackage{rotating}
\usepackage{amsmath}
\usepackage{bbm}
\usepackage{subfigure}
\usepackage{color}

\renewcommand{\v}[1]{{\bf #1}}
\newcommand{\intint}{\int\!\!\!\int}

\newcommand{\be}{\begin{equation}}
\newcommand{\ee}{\end{equation}}
\newcommand{\bea}{\begin{eqnarray}}
\newcommand{\eea}{\end{eqnarray}}

\begin{document}

\title{Discontinuity of the chemical potential in RDMFT for open-shell systems}
\author{N. Helbig}
\affiliation{Institut f\"ur Theoretische Physik, Freie Universit\"at Berlin, 
Arnimallee 14, D-14195 Berlin, Germany}
\affiliation{Unit\'e de Physico-Chimie et de Physique des Mat\'eriaux, 
Universit\'e Catholique de Louvain, B-1348 Louvain-la-Neuve, Belgium}
\affiliation{European Theoretical Spectroscopy Facility}
\author{N.N. Lathiotakis}
\affiliation{Institut f\"ur Theoretische Physik, Freie Universit\"at Berlin, 
Arnimallee 14, D-14195 Berlin, Germany}
\affiliation{Theoretical and Physical Chemistry Institute,
The National Hellenic Research Foundation,
Vass. Constantinou 48, 11635 Athens, Greece }
\affiliation{European Theoretical Spectroscopy Facility}
\author{E.K.U. Gross}
\affiliation{Institut f\"ur Theoretische Physik, Freie Universit\"at Berlin, 
Arnimallee 14, D-14195 Berlin, Germany}
\affiliation{European Theoretical Spectroscopy Facility}

\begin{abstract}
We employ reduced density-matrix functional theory in the calculation of the
fundamental gap of open-shell systems. The formula for the calculation of the
fundamental gap is derived with special attention to the spin of the neutral and
the charged systems. We discuss the effects of different functionals as well as the
changes due to different basis sets. Also, we investigate the importance of varying
the natural orbitals for the calculation of the fundamental gap.
\end{abstract}

\date{\today}

\maketitle

\section{Introduction}

Density functional theory (DFT) \cite{HK1964, KS1965} is a powerful tool to
calculate the electronic structure of atoms, molecules, and solids. Within DFT
observables are given as functionals of the particle density. In reduced
density-matrix functional theory (RDMFT) the 1-body reduced density matrix
(1-RDM) is used as the basic variable. RDMFT is based on Gilbert's theorem
\cite{G1975} which proves that each ground-state observable can, in principle,
be written as a functional of the 1-RDM. First-generation functionals
\cite{M1984, BB2002, GU1998} perform very well in the description of the
dissociation of small molecules. Second generation functionals were introduced
very recently \cite{GPB2005,PIR2006,PL2005} which improved both the performance
for small molecules \cite{GPB2005,PIR2006,PL2005} and also for the homogeneous
electron gas \cite{LHG2006}.

A key quantity in electronic structure calculations is
the band gap for semiconductors and insulators. It is defined as the difference
between the ionization potential $I$ and the electron affinity $A$ 
\be\label{delta}
\Delta = I-A,
\ee
where
\bea\label{eip}
I &=& E_{\mathrm{tot}}(N-1)-E_{\mathrm{tot}}(N)\,,\\
A &=& E_{\mathrm{tot}}(N)-E_{\mathrm{tot}}(N+1)\,.
\label{ea}
\eea
$E_{\mathrm{tot}}(N)$ denotes the ground-state energy of an $N$-electron system.
In the chemistry literature $\Delta/2$ is called the {\it chemical hardness} if
the system is finite. For simplicity we use the term {\it fundamental gap} for
both finite and extended systems throughout this article. We like to point out
that the fundamental gap differs from what is known as the optical gap. The
optical gap  is given as the energy necessary to excite the system from the
ground state to  the first excited state. Therefore, its size is reduced by the
binding energy of the created exciton compared to the fundamental gap.

Within density functional theory it can be shown \cite{PL1983,SS1985} that the 
fundamental gap is exactly given by
\be\label{deltadft}
\Delta=\Delta_{KS}+\Delta_{xc},
\ee
where $\Delta_{KS}$ is the energy difference between the lowest unoccupied and
the highest occupied Kohn-Sham states and $\Delta_{xc}$ is the discontinuity of
the exchange-correlation potential upon adding and subtracting a fractional
charge. This discontinuity is zero for LDA and GGA, so $\Delta_{KS}$ is the
prediction for the gap within these approximations. However, this prediction
deviates strongly from the experimental values. For semiconductors the
calculated gap underestimates the experimental value by typically 50\%. In
extreme cases, such as germanium, the gap vanishes within LDA.  Interestingly,
$\Delta_{KS}$ for the exact-exchange functional is very close to the
experimental gap for several systems \cite{SMVG1997,SMMVG1999}. Unfortunately, 
in the case of exact exchange $\Delta_{xc}$ is not zero and, in fact, was found
to be
much larger than $\Delta_{KS}$. Thus, if properly calculated, the band gaps
within exact exchange are highly overestimated compared to the experimental
values \cite{SMVG1997,SMMVG1999,SDD2005,GMR2006}. Exact exchange combined with
RPA correlation was recently shown to yield results very close to the
experimental values for Si, LiF, and solid Ar \cite{GMR2006} (provided the
discontinuity $\Delta_{xc}$ is properly included). Finally, a
recently introduced hybrid functional (HSE) \cite{HSE2003, HS2004} is reported
to give gaps in satisfactory agreement with experimental values for a set of 40
simple and binary semiconductors \cite{HPSM2005}. Especially, germanium is
predicted a semiconductor with a gap of 0.56~eV.

An alternative formula to (\ref{deltadft}) for the fundamental gap in DFT reads
\cite{PPLB1982}
\be\label{deltadftalt}
\Delta=\lim_{\eta\rightarrow 0^+}\left(\mu(N+\eta)-\mu(N-\eta)\right),
\ee
where $\mu$ is the chemical potential, and $N$ is the particle number of the
system. As Eq.~(\ref{deltadftalt}) suggests, the chemical potential has a
discontinuity at integer particle number $N$. In a recent 
paper~\cite{HLAG2006a}, we presented the analogous equation within reduced density-matrix
functional theory. In particular, we proved that the Lagrange multiplier used to
enforce the conservation of particle number is equal to the chemical potential. 
This theoretical development was applied to small finite and prototype periodic 
systems with very promising results. 
We like to emphasize that the analogy between DFT and RDMFT is not at all
trivial because of the $N$-representability condition in RDMFT. The occupation
numbers are restricted to the interval $[0,1]$ which leads to border minima. 
For this reason the generalization of the proof of Eq.~(\ref{deltadftalt})
from DFT to RDMFT is not straightforward.

In the present work, we deduce a relationship similar to Eq.~(\ref{deltadftalt})
for open-shell systems. The difficulty in generalizing Eq. (\ref{deltadftalt})
to the open-shell case arizes from the fact that adding/subracting a spin-up
electron to/from an open-shell ground state is not equivalent to
adding/subtracting a spin-down electron. Open-shell systems were recently
addressed in Ref.~\onlinecite{LHG2005} where it was demonstrated that  it is
reasonable to introduce two Lagrange multipliers to keep the number of 
electrons in each spin channel fixed seperately. An alternative description of
open-shell systems was introduced by Leiva and Piris~\cite{piris_os}. In that
desription, however, spin-up and spin-down occupations are equal for all
orbitals except the open-shell ones which are fully occupied by the majority
spin. The Lagrange multiplier is then spin independent. Here, we employ the 
treatment suggested in Ref.~\onlinecite{LHG2005} where each of the two Lagrange
multipliers is a function of the two particle numbers corresponding to the two
spin components. In the present work, these particle numbers are assumed to be
fractional. We show that a proper extension of Eq.~(\ref{deltadftalt}) is
possible with the resulting equation involving the discontinuities of both
Lagrange multipliers. The derivation is  presented in Section~\ref{sec_fundgap}.
Section \ref{sec_numres} contains results for a set of open-shell atoms and a
comparison of the closed- and open-shell treatment for systems where the neutral
system is actually closed-shell. We also investigate the performance of
different functionals in the calculation of  the fundamental gap.

\section{The fundamental gap in RDMFT}\label{sec_fundgap}

Reduced-density-matrix-functional theory (RDMFT) uses the one-body reduced
density matrix (1-RDM) 
\be
\gamma(\v x, \v x')=N\int\!d\v x_2...d\v x_N
\Psi^*(\v x', \v x_2,...\v x_N)\Psi(\v x, \v x_2,...\v x_N),
\ee
where $\Psi$ denotes the many-body wave function and $\v x=(\v r,\sigma)$. 
Integration over $d\v x$ means integration over space and 
summation over spin. Throughout this article we restrict ourselves, for
simplicity, to the "collinear" case where 
$\gamma(\v x, \v x')=\gamma(\v r\sigma,\v r'\sigma')$ is diagonal in spin space,
i.e.
\be
\gamma(\v r\sigma,\v r'\sigma')=\delta_{\sigma\sigma'}\gamma^\sigma(\v r,\v r').
\ee
By diagonalizing $\gamma^\sigma(\v r,\v r')$ one obtains the natural orbitals $\varphi_{j\sigma}$ and
the occupation numbers $n_{j\sigma}$, i.e.
\be
\gamma^\sigma(\v r,\v r')=\sum_{j=1}^\infty n_{j\sigma}
\varphi_{j\sigma}^*(\v r')\varphi_{j\sigma}(\v r).
\ee
To ensure the $N$-representability of $\gamma$ the occupation numbers are
restricted to the interval $[0,1]$ and sum up to the total number of particles
$N$.
In closed-shell systems the two spin-directions are identical, i.e.
\bea
n_{j\uparrow}&=&n_{j\downarrow},\\
\varphi_{j\uparrow}&=&\varphi_{j\downarrow}.
\eea

Within the spin-dependent formalism one can define spin-dependent electron 
affinities and ionization potentials by adding or removing an electron with 
specific spin
\bea
\label{isigma}
I^\sigma&=&E_{\mathrm{tot}}(N^\sigma-1,N^{\bar{\sigma}})-
E_{\mathrm{tot}}(N^\sigma,N^{\bar{\sigma}})\,, \\
\label{asigma}
A^\sigma&=&E_{\mathrm{tot}}(N^\sigma,N^{\bar{\sigma}})-
E_{\mathrm{tot}}(N^\sigma+1,N^{\bar{\sigma}})\,.
\eea
Here, $E_{\mathrm{tot}}(N^\sigma,N^{\bar{\sigma}})$ representes the ground-state
energy of a system with $N=N^\sigma+N^{\bar{\sigma}}$ electrons where $N^\sigma$
is the number of electrons with spin $\sigma$ and $N^{\bar{\sigma}}$ is the
number of electrons with the opposite spin, $\bar{\sigma}$.
Consequently, the ionization potential and electron affinity defined in 
Eq.~(\ref{ea}) are given by
\bea
I=\min_\sigma\{I^\uparrow,I^\downarrow\}\,, \\
A=\max_\sigma\{A^\uparrow,A^\downarrow\}\,,
\eea
i.e. they are respectively the smallest necessary energy for
taking away an electron and the maximum energy gained by adding an electron to
the neutral system.
The fundamental gap then reads
\be\label{defgap}
\Delta=\min_\sigma\{I^\uparrow,I^\downarrow\}
-\max_\sigma\{A^\uparrow,A^\downarrow\}\,.
\ee

In order to derive a formula analogous to Eq. (\ref{deltadftalt}) for the 
fundamental gap (\ref{delta}) within RDMFT we follow the
same path as in DFT \cite{SS1983, SS1985, K1986} and extend
the definition of the total-energy functional $E_{\mathrm{tot}}[\gamma]$
to systems with fractional particle number $M$. Throughout this paper 
we use the convention that $N$ denotes an integer number of particles
and $M$ a fractional. Such systems can be described as
an ensemble consisting of an $N$- and an $(N+1)$-particle state for $N\leq M\leq
N+1$. Let $\Psi_{N^\sigma,N^{\bar{\sigma}}}$ denote an $N$-particle wave
function with $N=N^\sigma+N^{\bar{\sigma}}$ where, as before, $N^\sigma$ is the number
of electrons with spin $\sigma$ and $N^{\bar{\sigma}}$ the number of
particles with the opposite spin, $\bar{\sigma}$. We consider an ensemble
where, compared to the charge-neutral $(N^\sigma,N^{\bar{\sigma}})$ system, the
number of spin-$\sigma$ particles is increased by $\eta^\sigma$. The statistical
operator describing such an ensemble is given by
\begin{multline}
\label{hatD}
\hat{D}_{N^\sigma+\eta^\sigma,N^{\bar{\sigma}}}=
(1-\eta^\sigma)\mid\!\Psi_{N^\sigma,N^{\bar{\sigma}}}
\rangle\langle\Psi_{N^\sigma,N^{\bar{\sigma}}}\!\mid\\
+\eta^\sigma\mid\!\Psi_{N^\sigma+1,N^{\bar\sigma}}
\rangle\langle\Psi_{N^\sigma+1,N^{\bar{\sigma}}}\!\mid\,.
\end{multline}
The expectation value of an operator $\hat{O}$ is then given by 
\be
O=tr\left(\hat{D}_{N^\sigma+\eta^\sigma,N^{\bar{\sigma}}}\hat{O}\right).
\ee
In particular, for $\hat{O}=\hat{\gamma}^{\sigma_1}(\v r,\v r')$, i.e. the
operator representing the 1-RDM of spin-$\sigma_1$ particles, we obtain
\begin{multline}
\label{defgammafrac}
\gamma^{\sigma_1}_{N^\sigma+\eta^\sigma,N^{\bar{\sigma}}}(\v r,\v r') = 
(1-\eta^\sigma)\gamma^{\sigma_1}_{N^\sigma,N^{\bar{\sigma}}}(\v r,\v r')\\
+\eta^\sigma\gamma_{N^\sigma+1,N^{\bar{\sigma}}}(\v r, \v r'),
\end{multline}
and for $\hat{O}=\hat{H}$, i.e. the Hamiltonian, we get the total ensemble energy
\begin{multline}
\label{etotm}
E_{\mathrm{tot}}(N^\sigma+\eta^\sigma,N^{\bar{\sigma}})=
(1-\eta^\sigma)E_{\mathrm{tot}}(N^\sigma,N^{\bar{\sigma}})\\
+\eta^\sigma E_{\mathrm{tot}}(N^\sigma+1,N^{\bar{\sigma}}).
\end{multline}
We note in passing that the ensemble weights in Eq. (\ref{hatD}) are such that
the correct normalization of spin-up and spin-down densities is achieved, i.e.
\bea
\int d^3r
\gamma^{\sigma}_{N^\sigma+\eta^\sigma,N^{\bar{\sigma}}}(\v r,\v r)&=&
N^\sigma+\eta^\sigma,\\
\int d^3r
\gamma^{\bar{\sigma}}_{N^\sigma+\eta^\sigma,N^{\bar{\sigma}}}(\v r,\v r)&=&
N^{\bar{\sigma}}.
\eea
Reformulating (\ref{etotm}) one obtains
\begin{multline}
E_{\mathrm{tot}}(M^\sigma,N^{\bar{\sigma}})=
E_{\mathrm{tot}}(N^\sigma,N^{\bar{\sigma}})\\
+\eta^\sigma
\left[E_{\mathrm{tot}}(N^\sigma+1,N^{\bar{\sigma}})-
E_{\mathrm{tot}}(N^\sigma,N^{\bar{\sigma}})\right]
\end{multline}
for $N^\sigma <M^\sigma=N^\sigma+\eta^\sigma <N^\sigma +1$. 
In analogy, for $N^\sigma -1<M^\sigma=N^\sigma-1+\eta^\sigma <N^\sigma $ 
the total energy is given by
\begin{multline}
E_{\mathrm{tot}}(M^\sigma,N^{\bar{\sigma}})=
E_{\mathrm{tot}}(N^\sigma-1,N^{\bar{\sigma}})\\
+\eta^\sigma
\left[E_{\mathrm{tot}}(N^\sigma,N^{\bar{\sigma}})-
E_{\mathrm{tot}}(N^\sigma-1,N^{\bar{\sigma}})\right].
\end{multline}
In other words, the total energy depends linearly on $\eta^\sigma$ with slope
$-A^\sigma$ for $N^\sigma\leq M^\sigma\leq N^\sigma+1$ and slope $-I^\sigma$ for 
$N^\sigma-1\leq M^\sigma\leq N^\sigma$.
Since $A^\sigma$ and $I^\sigma$ are in general not the same, the derivative
$\partial E_{\mathrm{tot}}(M^\sigma,N^{\bar{\sigma}})/\partial M^\sigma$
has a discontinuity at integer particle number $N^\sigma$. From Eqs.
(\ref{isigma})-(\ref{defgap}), one can conclude that the fundamental gap is
given by
\begin{multline}
\label{gap1}
\Delta=\min_\sigma\left\{\lim_{\eta^\sigma\to0^+}
\frac{\partial E_{\mathrm{tot}}(M^\uparrow,M^\downarrow)}{\partial
M^\sigma}\biggl|_{N^\sigma+\eta^\sigma,N^{\bar{\sigma}}}\right\}\\
-\max_\sigma\left\{\lim_{\eta^\sigma\to0^+}
\frac{\partial E_{\mathrm{tot}}(M^\uparrow,M^\downarrow)}{\partial
M^\sigma}\biggl|_{N^\sigma-\eta^\sigma,N^{\bar{\sigma}}}\right\}.
\end{multline}

In Ref~\onlinecite{LHG2005}, we argued that, for open-shell systems, the
following functional should be minimized 
\begin{multline}
\label{deff}
F[\gamma]=E_{\mathrm{tot}}[\gamma]\\-
\mu^\uparrow\left(\sum_{j=1}^\infty n_{j\uparrow}-M^\uparrow\right)-
\mu^\downarrow\left(\sum_{j=1}^\infty n_{j\downarrow}-M^\downarrow\right).
\end{multline}
The Lagrange multipliers $\mu^\uparrow$ and $\mu^\downarrow$ are introduced to
achieve given particle numbers $M^\uparrow$ and $M^\downarrow$. To prove the
formula for the fundamental gap we first show that these Lagrange multipliers
are nothing but the chemical potentials, i.e.
\be
\mu^\sigma(M_1^\uparrow,M_1^\downarrow)=\frac{\partial
E_{\mathrm{tot}}(M^\uparrow,M^\downarrow)}{\partial M^\sigma}
\biggl|_{M_1^\uparrow,M_1^\downarrow}.
\ee
The derivation of this formula differs significantly from the derivation of its
counterpart in DFT due to the above mentioned $N$-representability constraint. In
order for the 1-RDM to be connected to an anti-symmetric $N$-particle wave
function its occupation numbers have to be restricted to the interval [0,1]
\cite{C1963}. One can show that the same constraint ensures ensemble
$N$-representability for fractional particle number. As a result of this 
additional constraint, $\delta F/\delta\gamma$
need not vanish at the minimum energy. It is possible that certain occupation
numbers are pinned at the border of the interval while the true minimum is
obtained for values of $n_{j\sigma}$ outside this interval. The functional $F$
then has a border minimum, and therefore non-vanishing derivative, in all
directions where occupation numbers are pinned at zero or one.

We investigate the difference
\begin{multline}
E_{\mathrm{tot}}(M^\sigma+\eta^\sigma,M^{\bar{\sigma}})-
E_{\mathrm{tot}}(M^\sigma,M^{\bar{\sigma}})=\\
E\left[\gamma_{M^\sigma+\eta^\sigma,M^{\bar{\sigma}}}\right]
-E\left[\gamma_{M^\sigma,M^{\bar{\sigma}}}\right].
\end{multline}
A Taylor expansion of $E[\gamma_{M^\sigma+\eta^\sigma, M^{\bar{\sigma}}}]$ around 
$\gamma_{M^\sigma,M^{\bar{\sigma}}}$ yields
\begin{multline}\label{difference2}
E_{\mathrm{tot}}(M^\sigma+\eta^\sigma,M^{\bar{\sigma}})-
E_{\mathrm{tot}}(M^\sigma,M^{\bar{\sigma}})
=\\
\sum_{\sigma_1=\uparrow\downarrow}
\intint\! d^3 r d^3 r' \frac{\delta
E_{\mathrm{tot}}}{\delta\gamma^{\sigma_1}(\v r,\v r')}
\biggl|_{\gamma_{M^\sigma,M^{\bar{\sigma}}}^{\sigma_1}}\hspace*{3cm}\\
\times\left(\gamma_{M^\sigma+\eta^\sigma,M^{\bar{\sigma}}}^{\sigma_1}(\v r,\v r')-
\gamma_{M^\sigma,M^{\bar{\sigma}}}^{\sigma_1}(\v r,\v r')\right).
\end{multline}
For the functional derivative we employ (\ref{deff}) and obtain
\begin{multline}
\frac{\delta E_{\mathrm{tot}}}{\delta\gamma^{\sigma_1}(\v r, \v r')}=
\frac{\delta F}{\delta\gamma^{\sigma_1}(\v r, \v r')} \\ +
\mu^\uparrow\sum_{j=1}^\infty 
\frac{\delta n_{j\uparrow}}{\delta\gamma^{\sigma_1}(\v r, \v r')}+
\mu^\downarrow\sum_{j=1}^\infty 
\frac{\delta n_{j\downarrow}}{\delta\gamma^{\sigma_1}(\v r, \v r')}.
\end{multline}
The first term on the right is evaluated via the functional chain rule, i.e.
\begin{multline}
\frac{\delta E_{\mathrm{tot}}}{\delta\gamma^{\sigma_1}(\v r, \v r')}=
\sum_{\sigma=\uparrow\downarrow}\sum_{j=1}^\infty
\left(\frac{\delta F}{\delta n_{j\sigma}}+\mu^\sigma\right)
\frac{\delta n_{j\sigma}}{\delta\gamma^{\sigma_1}(\v r, \v r')}\\
+\int d^3 r''\left[
\frac{\delta F}{\delta\varphi_{j\sigma}(\v r'')}
\frac{\delta\varphi_{j\sigma}(\v r'')}{\delta\gamma^{\sigma_1}(\v r, \v r')}+ c.c\right].
\end{multline}
At the solution point, the variation with respect to the natural orbitals
vanishes such that the second term on the right is zero. The variation with
respect to the occupation numbers, however, need not vanish due to the
$N$-representability constraint. Equation (\ref{difference2}) therefore reduces
to
\begin{multline}
\label{difference3}
E_{\mathrm{tot}}(M^\sigma+\eta^\sigma,M^{\bar{\sigma}})-
E_{\mathrm{tot}}(M^\sigma,M^{\bar{\sigma}})
=\\
\sum_{\sigma_1=\uparrow\downarrow}\sum_p\intint d^3 r d^3 r'
\frac{\delta F}{\delta n_{p\sigma_1}}
\frac{\delta n_{p\sigma_1}}{\delta\gamma^{\sigma_1}(\v r, \v
r')}\Biggl|_{\gamma^{\sigma_1}_{M^\sigma,M^{\bar{\sigma}}}}\\
\times\left[\gamma_{M^\sigma+\eta^\sigma,M^{\bar{\sigma}}}^{\sigma_1}(\v r,\v r')-
\gamma_{M^\sigma,M^{\bar{\sigma}}}^{\sigma_1}(\v r,\v r')\right]\\ 
+\sum_{\sigma_1=\uparrow\downarrow}\sum_{j=1}^\infty\intint d^3 r d^3 r'
\mu^{\sigma_1}
\frac{\delta n_{j\sigma_1}}{\delta\gamma^{\sigma_1}(\v r, \v r')}
\Biggl|_{\gamma^{\sigma_1}_{M^\sigma,M^{\bar{\sigma}}}}\\
\times\left[\gamma_{M^\sigma+\eta^\sigma,M^{\bar{\sigma}}}^{\sigma_1}(\v r,\v r')-
\gamma_{M^\sigma,M^{\bar{\sigma}}}^{\sigma_1}(\v r,\v r')\right],
\end{multline}
where the first sum runs only over those occupation numbers pinned to the border of the
interval.
The variation of the occupation numbers can be calculated applying first order
perturbation theory to the eigenvalue equation of the 1-RDM
\be
\int d^3 r'\gamma^\sigma(\v r, \v r')\varphi_{j\sigma}(\v r')
=n_{j\sigma}\varphi_{j\sigma}(\v r)
\ee
which yields
\be
\frac{\delta n_{j\sigma}}{\delta \gamma^{\sigma}(\v r, \v r')}=
\varphi_{j\sigma}^*(\v r)\varphi_{j\sigma}(\v r').
\ee
In addition, we write the eigenvalues and eigenfunctions of
$\gamma^{\sigma_1}_{M^\sigma+\eta^\sigma,M^{\bar{\sigma}}}$ as
\be
\varphi^{M^\sigma+\eta^\sigma,M^{\bar{\sigma}}}_{j\sigma_1}=
\varphi_{j\sigma_1}+\delta\varphi_{j\sigma_1},
\quad
n_{j\sigma_1}^{M^\sigma+\eta^\sigma,M^{\bar{\sigma}}}=
n_{j\sigma_1}+\delta n_{j\sigma_1},
\ee
where $\varphi_{j\sigma_1}$ and $n_{j\sigma_1}$ denote the natural orbitals and
occupation numbers of $\gamma^{\sigma_1}_{M^\sigma,M^{\bar{\sigma}}}$. 
Equation (\ref{difference3}) then reduces to
\begin{widetext}
\bea\nonumber
E_{\mathrm{tot}}(M^\sigma+\eta^\sigma,M^{\bar{\sigma}})-
E_{\mathrm{tot}}(M^\sigma,M^{\bar{\sigma}})
&=&\!\sum_{\sigma_1=\uparrow\downarrow}\sum_p
\frac{\delta F}{\delta n_{p\sigma_1}}
\left(\delta n_{p\sigma_1}+
\int d^3 r\: n_{p\sigma_1}
\left[\varphi_{p\sigma_1}(\v r)\delta \varphi_{p\sigma_1}^*(\v
r)+\varphi_{p\sigma_1}^*\delta\varphi_{p\sigma_1}(\v r)\right]\right)\\
&+&\!\sum_{\sigma_1=\uparrow\downarrow}\sum_{j=1}^\infty
\mu^{\sigma_1}
\left(\delta n_{j\sigma_1}+\int\! d^3 r\: n_{j\sigma_1}
\left[\varphi_{j\sigma_1}(\v r)\delta \varphi_{j\sigma_1}^*(\v
r)+\varphi_{j\sigma_1}^*\delta\varphi_{j\sigma_1}(\v r)\right]\right),
\label{energydiff}
\eea
\end{widetext}
where we only kept terms up to first order. In order for the natural
orbitals to remain normalized the changes have to be orthogonal to the original
orbitals, i.e.
\be\label{normal}
\int d^3 r\:\varphi_{j\sigma}^*(\v r)\delta\varphi_{j\sigma}(\v r)=0.
\ee 
Therefore, the integrals one the right-hand-side of Eq. (\ref{energydiff}) 
vanish. The sum over all changes in the occupation numbers
has to give $\eta^\sigma$ in order for the new occupation numbers to sum up to
the correct particle number. Hence, we obtain
\begin{multline}
E_{\mathrm{tot}}(M^\sigma+\eta^\sigma,M^{\bar{\sigma}})-
E_{\mathrm{tot}}(M^\sigma,M^{\bar{\sigma}})=\\
\mu^\sigma\eta^\sigma+\sum_{\sigma_1=\uparrow\downarrow}\sum_p
\frac{\delta F}{\delta n_{p\sigma_1}}\delta n_{p\sigma_1}
\end{multline}
Finally, we discuss the contribution of the pinned states. As stated before, for
these states $\delta F/\delta n_p$ is different from zero and the true minimum
of the functional lies outside the interval [0,1]. More specifically, it lies
at a finite distance from the border of the interval such that the addition or
subtraction of an infinitesimal fraction $\eta^\sigma$ of a particle cannot move
the minimum into the interval. Therefore, these particle numbers remain pinned
upon adding or subtracting an infinitesimal $\eta^\sigma$, i.e. 
$\delta n_{p\sigma_1}$ is zero in
the limit $\eta^\sigma\rightarrow 0$. We therefore conclude
\bea\nonumber
\mu^\sigma(M_1^\uparrow,M_1^{\downarrow})&=&\\
\nonumber
&&\hspace*{-2.5cm}\lim_{\eta^\sigma\rightarrow 0^+}\!\left(
\frac{E_{\mathrm{tot}}(M^\sigma+\eta^\sigma,M^{\bar{\sigma}})-
E_{\mathrm{tot}}(M^\sigma,M^{\bar{\sigma}})}{\eta^\sigma}
\right)\!\biggl|_{M_1^\uparrow,M_1^{\downarrow}}\\
&=&\frac{\partial E_{\mathrm{tot}}(M^\uparrow,M^\downarrow)}{\partial M^\sigma}
\biggl|_{M_1^\uparrow,M_1^{\downarrow}}.
\eea
Using Equation (\ref{gap1}) we obtain the final result for the fundamental gap
\begin{multline}
\label{gap}
\Delta=\min_{\sigma}\left(\lim_{\eta^\sigma\to 0^+}
\mu^\sigma(M^\sigma+\eta^\sigma, M^{\bar{\sigma}})\right)\\
-\max_{\sigma}\left(\lim_{\eta^\sigma\to 0^+}
\mu^\sigma(M^\sigma-\eta^\sigma, M^{\bar{\sigma}})\right).
\end{multline}
The derivation of Eq.~(\ref{gap}) concerns the exact
exchange-correlation energy functional of the 1-RDM. Since only approximations
are available, the question is whether Eq.~(\ref{gap}) is still
useful. This question is the main subject of the next section.
 
In Ref.~\onlinecite{HLAG2006a}, a single, spin independent $\mu$ (for
closed-shell systems) was shown to have a discontinuity as a function of a
fractional total number of electrons  which is equally distributed in the two
spin channels. The application of that theory to an open-shell system would
give  the spin resolved $\mu^\sigma$ as a function of a unique $M$.  In the
present work, we add/subtract a fractional part of  an electron to/from a
specific spin channel. Consequently, the system becomes an open-shell system
even if the neutral system is closed-shell.  Thus, we have four functions
$\mu^\uparrow(M^\uparrow,N^\downarrow)$, 
$\mu^\downarrow(N^\uparrow,M^\downarrow)$,
$\mu^\uparrow(N^\uparrow,M^\downarrow)$, and
$\mu^\downarrow(M^\uparrow,N^\downarrow)$, where $N^\uparrow$, $N^\downarrow$ 
are fixed to the integer values of the neutral system. Of these four, only the
first two show a discontinuity. The correct gap is then given by
Eq.~(\ref{gap}), where the min and max functions take care of the  selection of
the smallest ionization potential and the largest  electron affinity.
Alternatively, one can employ Eqs. (\ref{delta})-(\ref{ea}) for the calculation
of the fundamental gap. Both approaches are exact, in the sense that, given the 
exact functional of $\gamma$, they both reproduce the  fundamental gap.  It is
interesting to see if they give the same numbers for approximate functionals as
well. This is also one of the questions we address in the next section.

To answer the above questions, one needs to minimize the  approximate
functionals for fractional number of particles to get 
$\mu^{\uparrow,\downarrow}(M^\sigma,N^{\bar{\sigma}})$. The extension of the minimization
procedure to fractional particle  numbers, which is in complete accordance with
the proof  we presented above, requires us to perform the minimization in the 
domain of $\gamma^{\sigma_1}_{N^\sigma+\eta^\sigma,N^{\bar{\sigma}}}$, which
are given by Eq. (\ref{defgammafrac}). In principle one then has to minimize
the total energy with respect to
$\gamma^{\sigma_1}_{N^\sigma,N^{\bar{\sigma}}}$ and
$\gamma^{\sigma_1}_{N^\sigma+1,N^{\bar{\sigma}}}$ under the known
$N$-representability constraints that their occupation numbers are between 0
and 1 and sum up to the correct particle numbers. However, this procedure,
involving the density matrices for $N$ and $N+1$ particles, is not very
practical. On the contrary, it is desirable to minimize  with respect to
$\gamma^{\sigma_1}_{N^\sigma+\eta^\sigma,N^{\bar{\sigma}}}$ directly under the
appropriate constraints. We prove elsewhere~\cite{HLSG2006} that the
appropriate constraints for such a minimization are
\begin{equation}
 0\leq n^{(M^\sigma,M^{\bar{\sigma}})}_{j\sigma_1} \leq 1 \quad \forall j, \quad
 \sum_j n^{(M^\sigma,M^{\bar{\sigma}})}_{j\sigma_1}=M^{\sigma_1}\,.
\label{nconstr}
\end{equation}
In other words, the domain of 
$\gamma^{\sigma_1}_{N^\sigma+\eta^\sigma,N^{\bar{\sigma}}}$ which can be
represented as the weighted average Eq.~(\ref{defgammafrac}) is identical to
the domain  of  $\gamma^{\sigma_1}_{N^\sigma+\eta^\sigma,N^{\bar{\sigma}}}$
whose eigenvalues satisfy Eq.~(\ref{nconstr}). The above statement is quite
significant since the constraint of Eq.~(\ref{nconstr}) is much simpler
and completely analogous to the case of integer particle numbers. The
implementation is therefore  a rather simple extension of the case of integer
particle numbers.

\section{Numerical results}\label{sec_numres}

In this section, we study the behavior of $\mu$ as a function of the 
fractional particle number for some atoms and molecules using approximate
functionals of the 1-RDM. Our aim is to investigate whether there exists  a
discontinuity in $\mu(M)$ and how it compares to the fundamental gap.   The
implementation we used for finite systems can be applied to both closed- and
open-shell~\cite{LHG2005} configurations. Some results for closed-shell systems
were presented in  Ref.~\onlinecite{HLAG2006a}. Here, we give an extended
analysis for both closed- and open-shell systems.

For the open-shell treatment, we use the extension of the functional of 
Goedecker-Umrigar\cite{GU1998} described in Ref.~\onlinecite{LHG2005}. We also
investigate whether other functionals reproduce a discontinuity in a
closed-shell treatment. For this purpose we consider the functionals of 
Piris\cite{PIR2006,PL2005}, where the self-interaction (SI) terms are 
explicitly removed, and the M\"uller functional and the most recent BBC 
functionals of Gritsenko et al\cite{GPB2005} which contain self-interaction
terms.

The implementation is based on the GAMESS  program\cite{SBBEGJKMNSWDM1993}
which we use for the calculation  of the one and two-electron integrals. The
minimization with respect to the occupation numbers and natural orbitals is
then performed using the conjugate gradient method. Our program treats both
closed- as well as open-shell systems using the restricted open-shell
RDMFT~\cite{LHG2005}. In short, we assume spin-dependent occupation numbers
(and chemical potentials) but spin-independent natural orbitals. In that way,
our method is  in complete analogy to spin restricted open-shell Hartree-Fock.

\begin{figure}
\subfigure{\includegraphics[width=0.45\textwidth, clip]{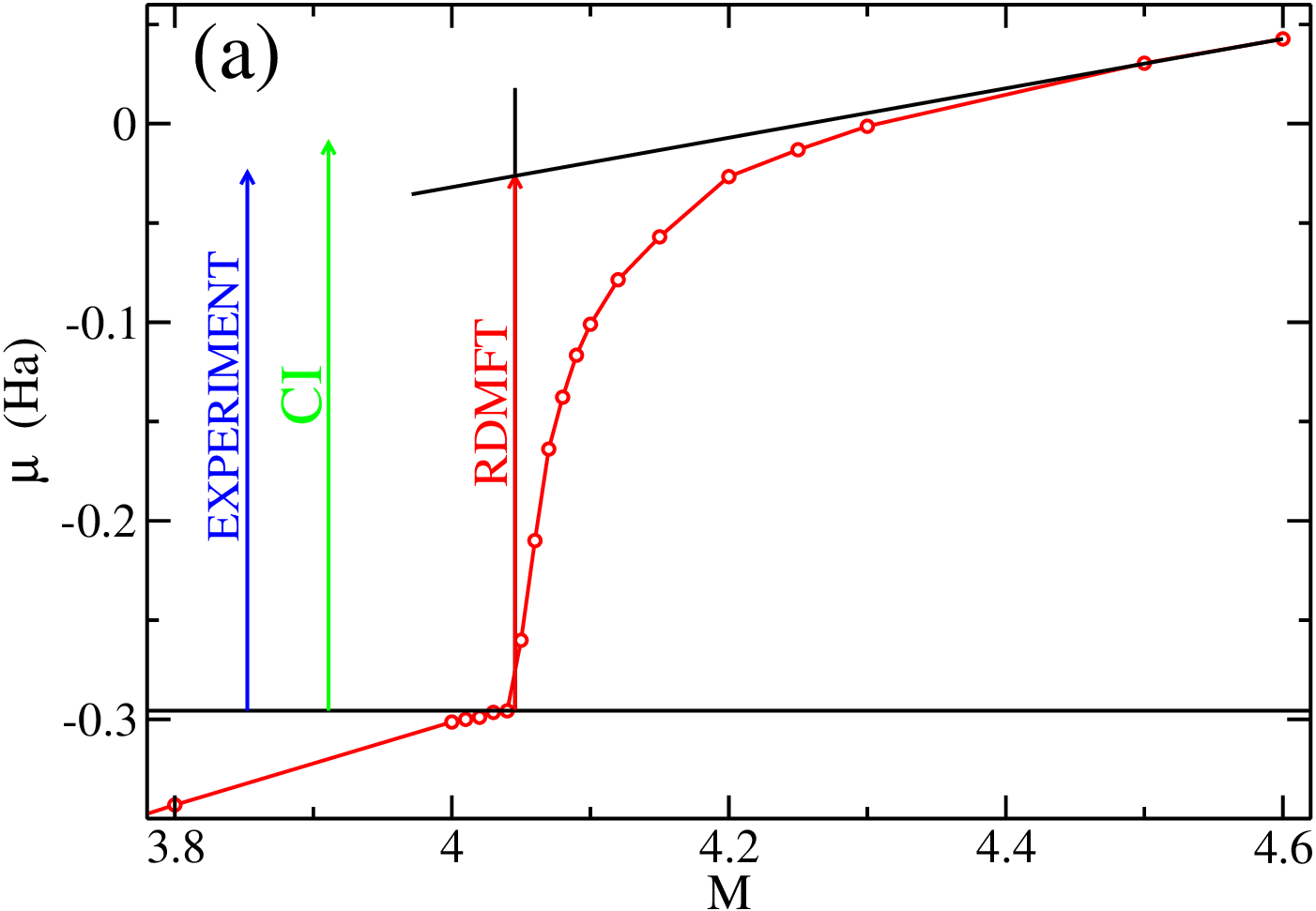}}
\subfigure{\includegraphics[width=0.45\textwidth, clip]{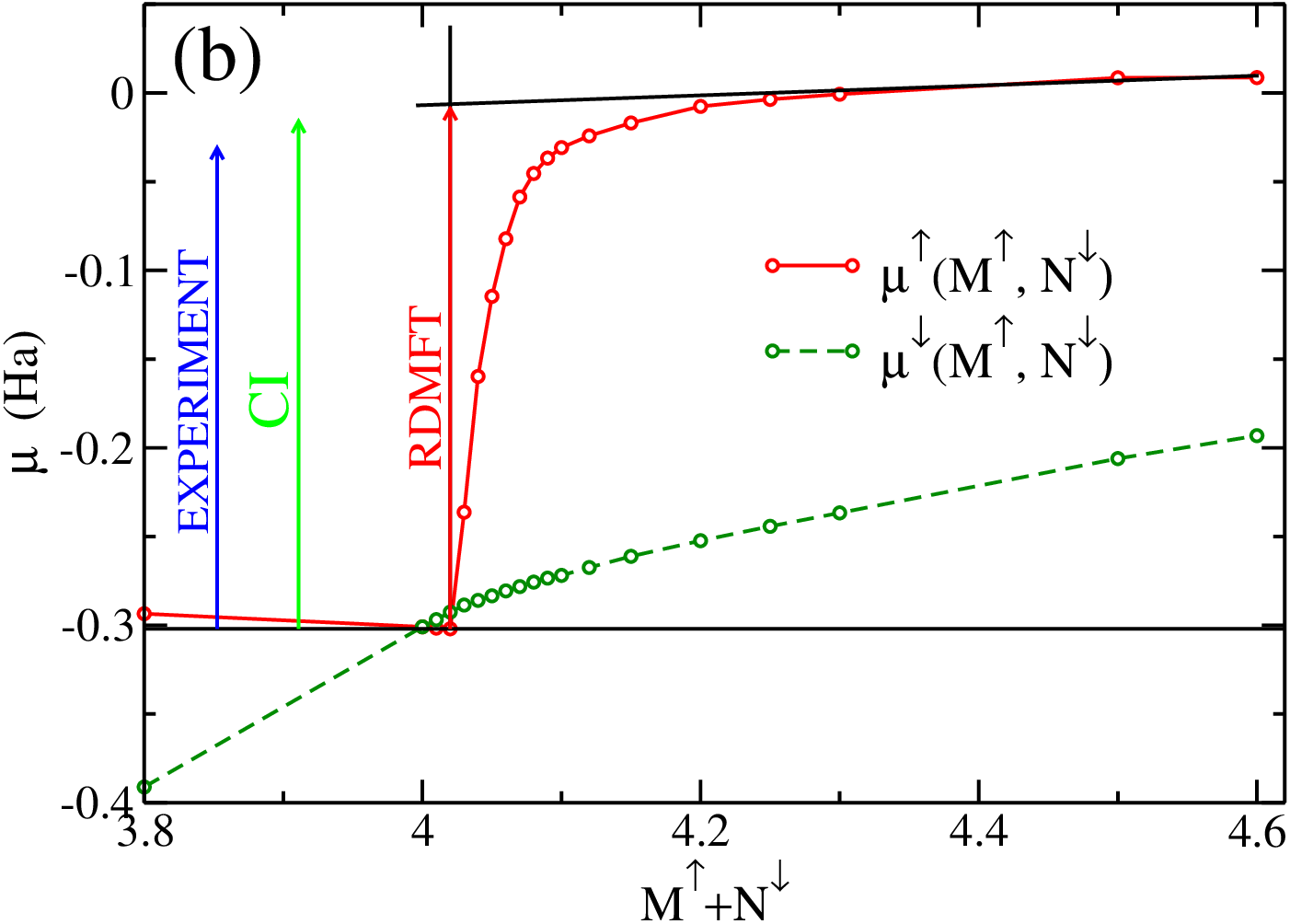}}
\caption{\label{fig:mu_N_LiH_mol} The behavior of $\mu$ as a function
of a fractional electron number $M$ for the  LiH  molecule in a closed-shell
treatment (a) and $\mu^{\uparrow,\downarrow}(M^\uparrow, N^\downarrow)$ for an open-shell
treatment (b). For 
comparison, the experimental and CI values of the fundamental gap are
included.}
\end{figure}

In Fig.~\ref{fig:mu_N_LiH_mol}a, we show $\mu(M)$ for the LiH molecule using 
the GU functional in the closed-shell treatment, i.e. the extra charge is 
equally distributed over the two spin channels. Fig.~\ref{fig:mu_N_LiH_mol}b
shows $\mu^\uparrow(M^\uparrow, N^\downarrow)$ and  $\mu^\downarrow(M^\uparrow,
N^\downarrow)$ for  the open-shell treatment of the LiH molecule, using again
the GU functional. In the open-shell treatment the additional charge is
exclusively added to one spin channel, and here we choose the spin-up channel.
Clearly,  $\mu(M)$  in Fig.~\ref{fig:mu_N_LiH_mol}a and 
$\mu^\uparrow(M^\uparrow, N^\downarrow)$ in Fig.~\ref{fig:mu_N_LiH_mol}b show a
pronounced step which resembles the  discontinuity that one expects for the
exact functional. This step has two important features: the first is that it
occurs not exactly at $M=4$, i.e. the exact, integer number of electrons. It is
rather shifted slightly  to the right. The shift is of the order of 0.05 of an
electron in  Fig.~\ref{fig:mu_N_LiH_mol}a and is reduced to 0.02 in 
Fig.~\ref{fig:mu_N_LiH_mol}b. A closer  look at the solution reveals that the
bottom of the step appears  exactly at the point where the occupation number of
the HOMO gets equal to one. After that point it has to remain one due to the
$N$-representability constraints, Eq.~(\ref{nconstr}). The pinning of the
occupation number  of the HOMO to one results in the rapid increase of $\mu$. 
Since adding charge to one spin channel only results in faster pinning of the
HOMO state it is not surprising that the step in the open-shell treatment is 
shifted to the left. Upon increasing the extra charge further, $\mu$ is a smooth
function, i.e. the upper edge of the step is rounded off. In the closed-shell
treatment $\mu(M)$ shows a linear dependence outside the step region, which is
significantly reduced in  $\mu^\uparrow$  in closer resemblance to the exact
behavior. A more detailed investigation reveals that the slope of $\mu(M)$ is
the average of the slopes  of $\mu^\uparrow(M^\uparrow, N^\downarrow)$ and
$\mu^\downarrow(M^\uparrow, N^\downarrow)$. To extract a value for the
discontinuity we use a backwards projection as  shown in
Fig.~\ref{fig:mu_N_LiH_mol}. This method reduces to the exact discontinuity if
$\mu$ is a true step function. The extracted values, as well as the gaps of
other finite systems are given in Table~\ref{tab_gap}. We should also keep in
mind  that DFT methods like LDA and GGA underestimate the gap by typically 50\%.
Although the procedure of backwards projection might seem rather crude and
arbitrary, we should mention that the agreement with experiment is rather
satisfactory for both close- and open-shell treatment. As one can see, for LiH,
the quantitative agreement is slightly better for a closed-shell treatment.
Nevertheless, the open-shell treatment should be prefered because $\mu$ then
resembles the exact step function much closer making the backward projection
less ambigious. 

\begin{figure}
\includegraphics[width=0.45\textwidth, clip]{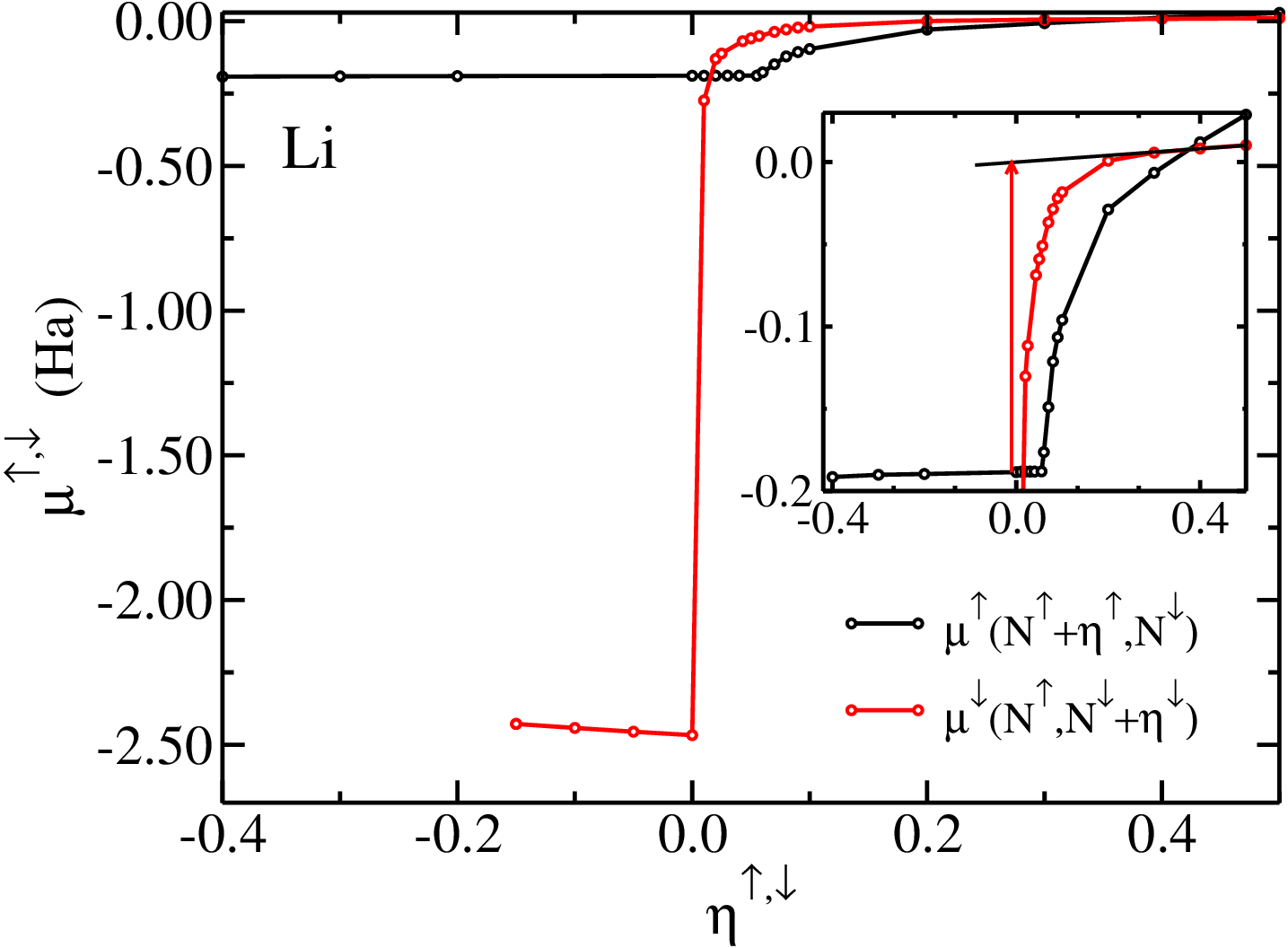} 
\caption{\label{fig:mu_N_Li_at} The behavior of $\mu^\sigma$ as a function
of an electron fraction $\eta^\sigma$ added (subtracted) to the 
neutral system for the Li atom. In the inset, we show an enlargement of the 
region where we extract the value for the gap from the difference of 
the upper level of $\mu^\downarrow(N^\uparrow, M^\downarrow)$ and 
the lower level of $\mu^\uparrow(M^\uparrow, N^\downarrow)$.}
\end{figure}

\begin{figure}
\subfigure{\includegraphics[width=0.45\textwidth, clip]{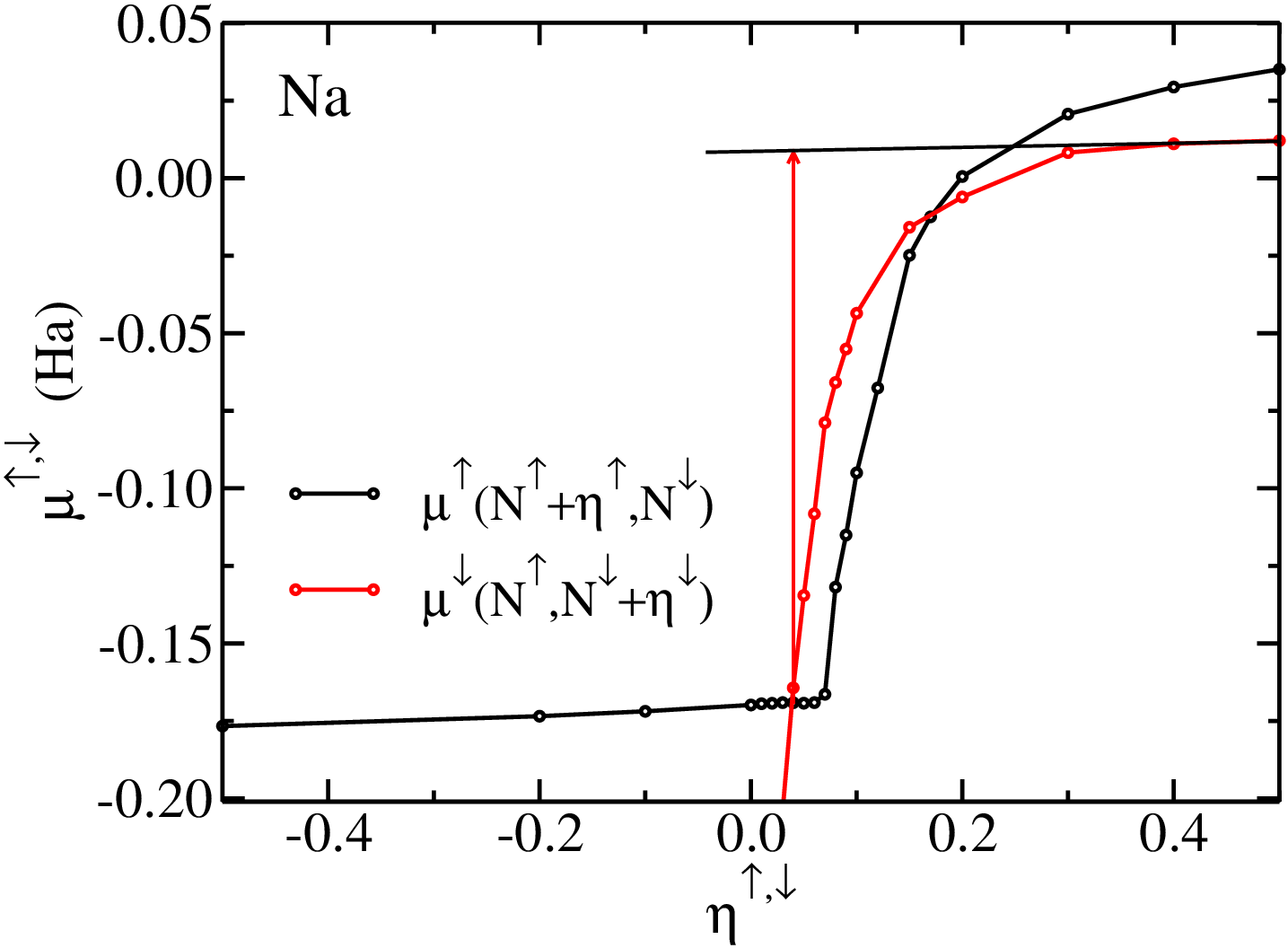}}
\subfigure{\includegraphics[width=0.45\textwidth, clip]{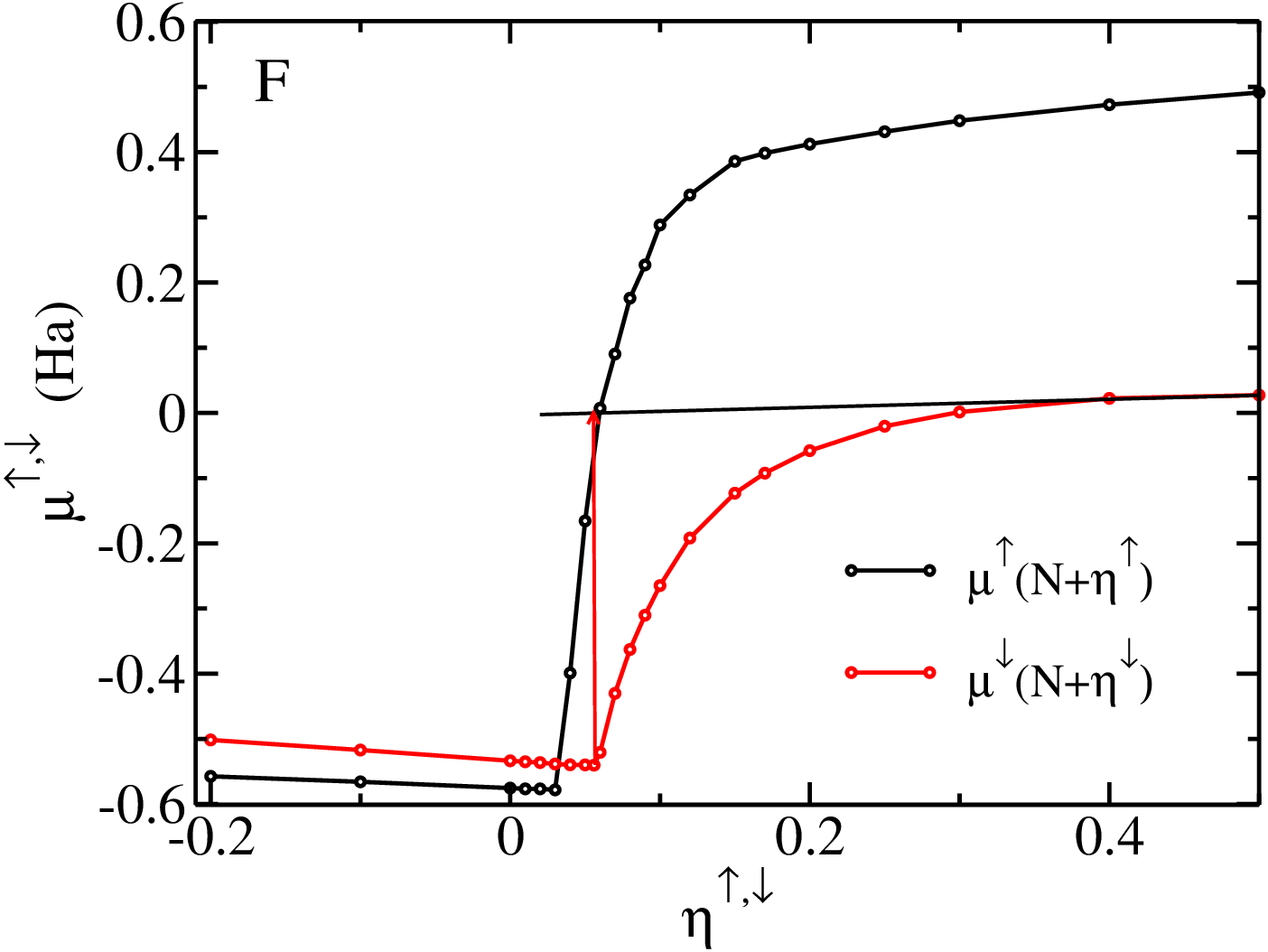}}
\caption{\label{fig:mu_N_Na_F} The behavior of $\mu^\sigma$ as a function
of an electron fraction $\eta^\sigma$ added (subtracted) to the 
neutral system for Na and F atoms. For Na, we show only the regions where 
the values of the gap are extracted from.}
\end{figure}

For open-shell systems, varying $M^\uparrow$ or $M^\downarrow$ is not
equivalent anymore. Thus, we can study the behavior of both $\mu^\uparrow$ and
$\mu^\downarrow$  as functions of $M^\uparrow$ or  $M^\downarrow$. We
investigate the open-shell atoms Li, Na, and F varying $M^\uparrow$ or
$M^\downarrow$ away from the neutral configurations. In the following, we use
the convention that spin up is always the majority spin channel. In
Fig.~\ref{fig:mu_N_Li_at}, we show the results for $\mu^\sigma$  for the Li
atom. Only the chemical potential corresponding to the spin direction whose
particle number is changed shows a discontinuity as already observed for the
LiH molecule. Therefore, we only plot $\mu^\uparrow(M^\uparrow, N^\downarrow)$
and $\mu^\downarrow(N^\uparrow, M^\downarrow)$. Again, pronounced steps
resembling the discontinuity  of the exact theory are present.  The prediction
for the gap is then selected using  Eq.~(\ref{gap}) and the backwards
extrapolation procedure described earlier. The values obtained for the gaps are
listed in Table~\ref{tab_gap}. According to  Eq.~(\ref{gap}), the gap for the
Li atom is given by the difference between the backwards projected upper part
of $\mu^\downarrow(N^\uparrow,M^\downarrow)$ and the lower part of
$\mu^\uparrow(M^\uparrow, N^\downarrow)$.  In Fig.~\ref{fig:mu_N_Na_F}, we show
the analogous results for the Na and F atoms. The picture for the Na atom is
very similar to Li.  On the other hand, for the F atom, the gap is given by 
$\mu^\downarrow(N^\uparrow,M^\downarrow)$ alone.  It is interesting that the
position of the upper and lower parts of the  $\mu^\sigma$ corresponds to the
actual process of adding and removing  electrons to the system. Thus, for Li
and Na atoms,  it is favorable to remove  an electron from the majority spin
channel (up) and add an extra electron to the minority spin channel (down). As
a consequence the gap is given by the difference between the upper part of
$\mu^\downarrow(N^\uparrow,M^\downarrow)$ and the  lower part of
$\mu^\uparrow(M^\uparrow,N^\downarrow)$. For a F atom, on the other hand, it is
favorable to add an electron to, or remove from, the minority spin channel.
Thus, the gap is given by $\mu^\downarrow(N^\uparrow, M^\downarrow)$ alone.

\begin{table}
\begin{center}
\begin{tabular}{lcccc} \hline\hline
System & RDMFT & RDMFT & Other & Experiment \\ 
       & $\mu(M)$ step & Eqs.~(\ref{delta})-(\ref{ea}) & theoretical &  \\ \hline
Li  &  0.18 & 0.202 &  0.175\footnotemark[1]  & 0.175\footnotemark[2]  \\
Na  &  0.18 & 0.198  &  0.169\footnotemark[3]  & 0.169\footnotemark[2]  \\
F   &  0.54 & 0.549 &  &0.514\footnotemark[2]\\
LiH &  0.27\footnotemark[4],0.29\footnotemark[5] & 0.271 &  0.286\footnotemark[6] &
0.271\footnotemark[7]  \\ 
\hline\hline
\end{tabular}
\end{center}
\caption[The fundamental gap of atoms and small molecules]
{\label{tab_gap} The prediction for the fundamental gap for several atoms and
small molecules
using the size of the step of $\mu(M)$, and a direct calculation through 
Eqs.~(\ref{delta})-(\ref{ea}) 
for the GU functional compared with experimental and other theoretical values.
For the direct application of the Eqs.~(\ref{delta})-(\ref{ea}),
the total energies of the positive and negative
ions were calculated.

\footnotemark[1] QCI from Ref. \cite{MOP1994}

\footnotemark[2] from Ref. \cite{RS1985}

\footnotemark[3] Ionization potential from \cite{MOP1994}, electron affinity
from \cite{GM2004}

\footnotemark[4] Closed-shell treatment

\footnotemark[5] Open-shell treatment

\footnotemark[6] CISD using the same basis set as in RDMFT

\footnotemark[7] Ionization potential from \cite{IW1975}, electron affinity from
\cite{SG2000}}
\end{table}

In Table \ref{tab_gap} we give the results obtained by the backward
extrapolation for the systems discussed in this paper. As one can see, they
agree very well with experimental values for the fundamental gap as well as
other theoretical calculations. For finite systems, one can also calculate the
gap by performing three  total energy calculations, for the $N$, the $N+1$ and
$N-1$ particle systems and use Eqs.~(\ref{delta}-\ref{ea}). The values for the
gap obtained in this way are given in Table \ref{tab_gap} for comparison. One
should keep in mind that for solid state systems, this procedure does not apply
because the addition or the removal of a single electron to an infinite solid is
meaningless. For such systems, the recipe introduced in this work is expected to
be valuable.

\begin{figure}
\includegraphics[width=0.45\textwidth, clip]{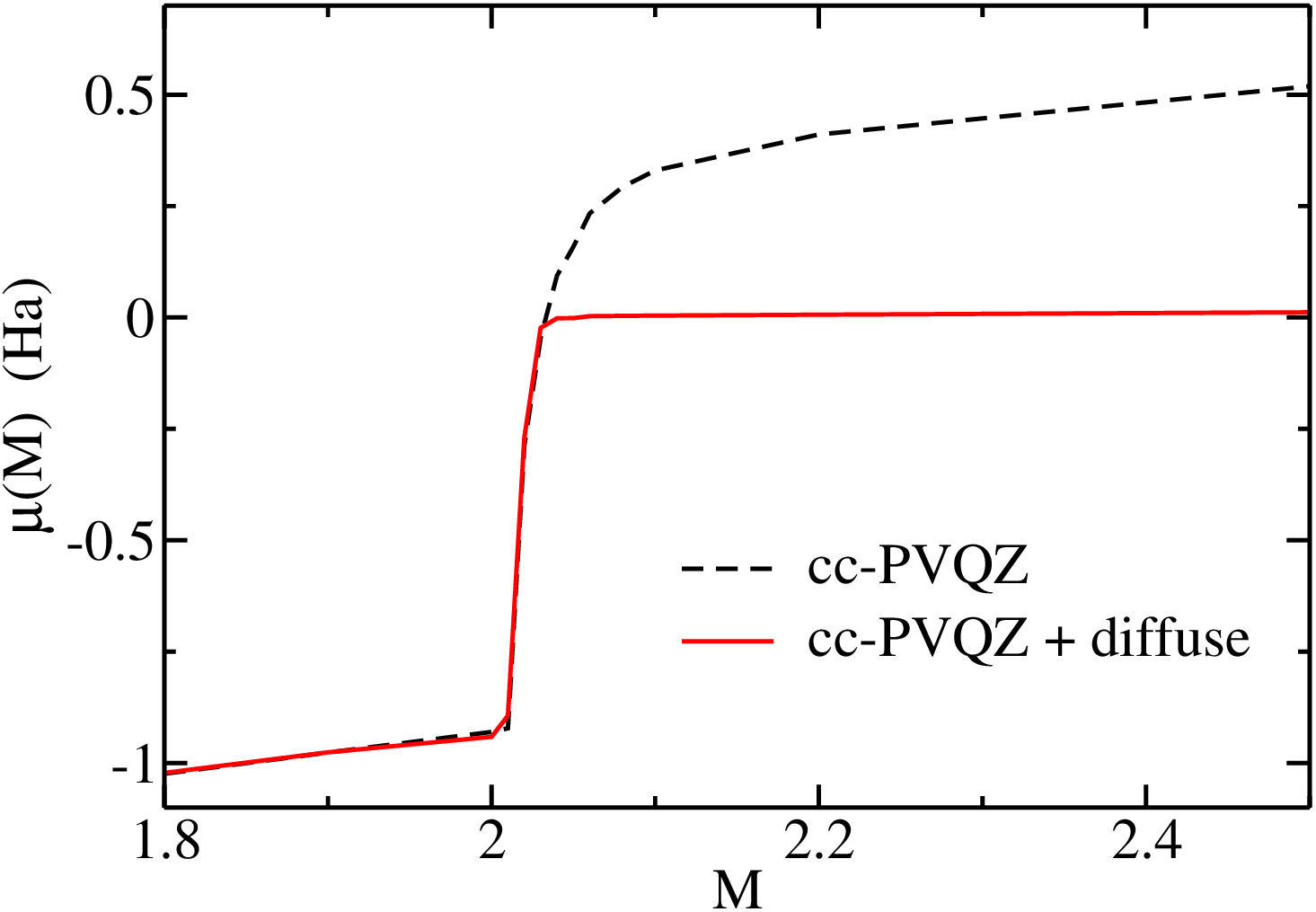}
\caption{\label{fig:mu_N_He_at} The function $\mu(M)$ for
the He atom using the cc-PVQZ basis set without and with an
additional very diffuse s-type basis function.}
\end{figure}

Of course the question arises whether the system with excess charge is
correctly described by the basis set we used. Usually, atomic basis sets  are
optimized to correctly describe the neutral system resulting in basis functions
which are all localized.  Therefore, the charged system might be predicted to
have a localized bound state despite the fact that the configuration of a
neutral atom and a free completely delocalized electron is energetically
favorable.  A prominent example of a system not having a negative ion is the
He-atom.  We study the behavior of  $\mu(M)$ with two different basis-sets: the
CC-PVQZ basis-set  and CC-PVQZ enlarged by a very diffuse s-type function. As
one can see in Fig.~\ref{fig:mu_N_He_at}, the state of the additional
fractional electron  is better described by the enlarged basis set. In this
case, the electron affinity is zero and the gap is given by the IP alone. 
Interestingly, the inclusion of a diffuse function leads to a sharper step of
$\mu(M)$  in close resemblance to the discontinuity of the exact functional. We
also add extra  diffuse functions in the basis-sets of both Li and H in the
calculation of $\mu(M)$ for the LiH molecule. We do not observe any effect on
$\mu(M)$, which is a clear evidence for the fact  that LiH binds an extra
electron and that the localized basis-set  is appropriate for describing the
state of the charged system. 

\begin{figure}
\subfigure{\includegraphics[width=0.45\textwidth, clip]{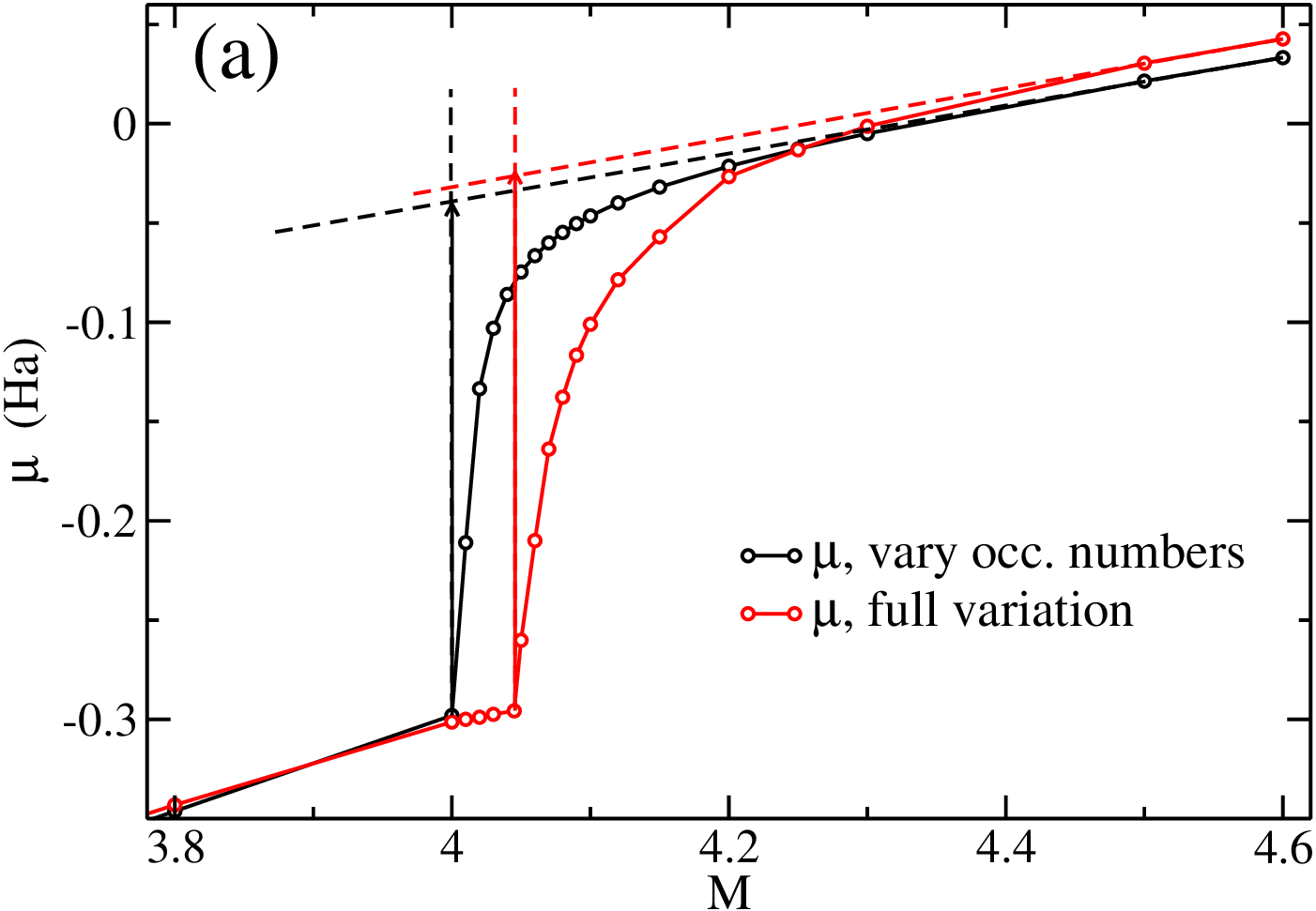}}
\subfigure{\includegraphics[width=0.45\textwidth, clip]{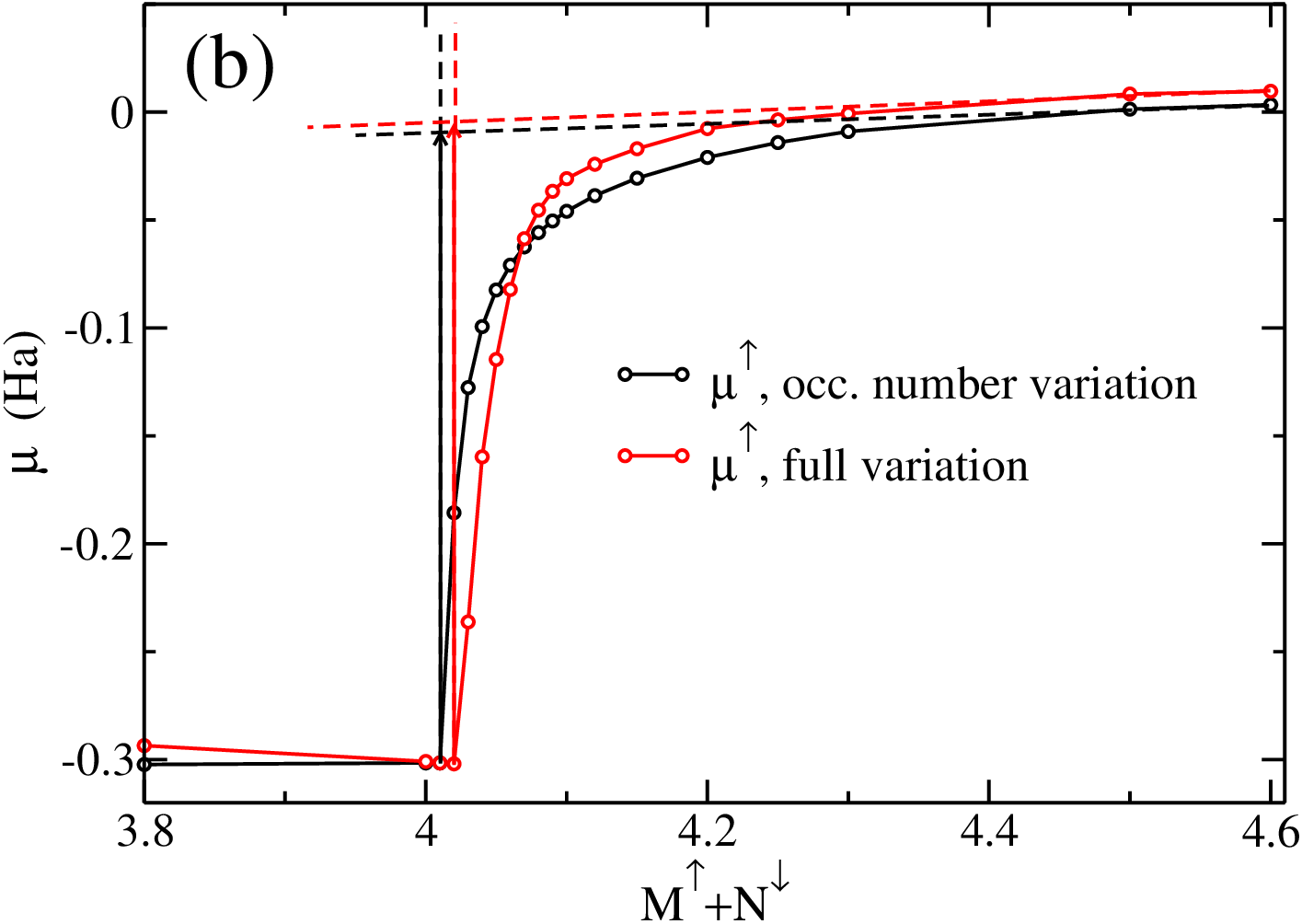}}
\caption{\label{fig:mu_N_noorb} The function $\mu(M)$ ($\mu^\uparrow(M^\uparrow, N^\downarrow)$)
for the LiH molecule using the closed-shell (a) and the open-shell treatment (b), 
with occupation number variation (using the Hartree
Fock orbitals) and with full variation (both occupation numbers and the orbitals).}
\end{figure}

In order to investigate the importance of the variation of the natural orbitals
for the discontinuity of $\mu$ we perform, apart from the full variation
described so far, a calculation where only the occupation numbers are optimized
while for the natural orbitals we keep the initial Hartree-Fock orbitals. In
Fig.~\ref{fig:mu_N_noorb} we compare these two procedures for both a closed- and
an open-shell calculation. As one can see from the plots, the main contribution
to the discontinuity arises from the variation of the occupation numbers. In the
closed-shell calculation we obtain a discontinuity of 0.27~Ha for the full
variation compared to 0.26~Ha if we vary the occupation numbers only. In other
words, only about $4\%$ of the discontinuity are due to the optimization of the
natural orbitals. This picture remains unchanged if we use the open-shell
procedure where we obtain 0.31~Ha for the full variation and 0.29~Ha for the
variation of the occupation numbers alone. 

\begin{figure}
\includegraphics[width=0.45\textwidth, clip]{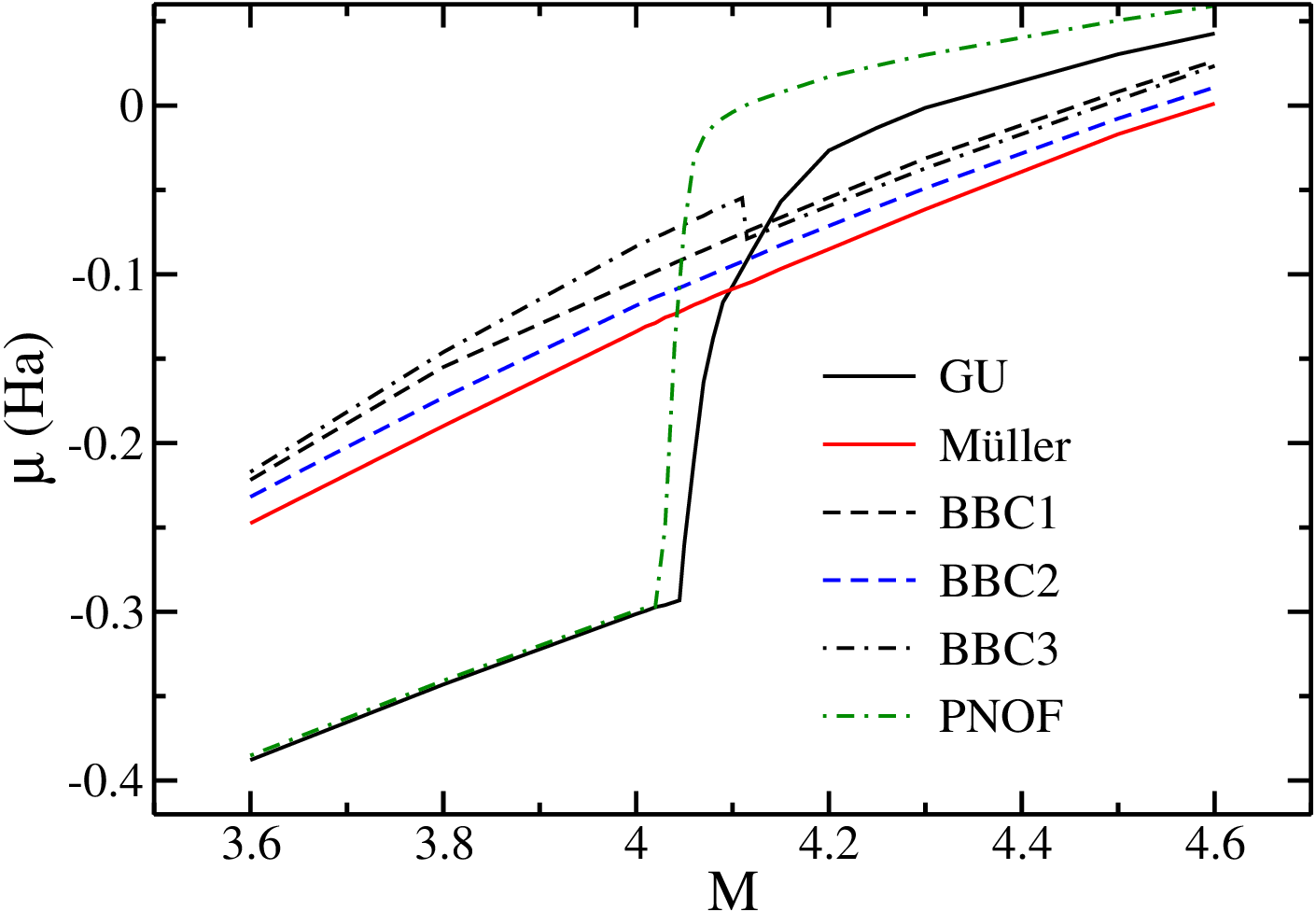} 
\caption{\label{fig:all_funct} The function $\mu(M)$ 
for the LiH molecule using the closed-shell treatment
for the Goedecker-Umrigar, the M\"uller, BBC1, BBC2, BBC3, and
PNOF functionals. The first and the last involve a complete
removal of the SI terms. Only these two reproduce a
pronounced step in resemblance to the discontinuity of the exact theory.}
\end{figure}

In all the calculations presented so far, we have used the functional of
Goedecker and Umrigar, which involves the complete removal of the
self-interaction terms.  It is interesting to study the behavior of $\mu(M)$
using different functionals, like for instance the recent BBC functionals of
Gritsenko et al~\cite{GPB2005} and the  PNOF of Piris\cite{PIR2006,PL2005}. In
the BBC1 and BBC2 functionals, the SI terms are present while in the BBC3, they
are partially removed. However, the SI terms for the bonding and the
anti-bonding orbitals remain. In the  PNOF they are fully removed as in GU. In
Fig.~\ref{fig:all_funct}, we plot $\mu(M)$ for LiH using the closed-shell
treatment, for all these functionals. Surprisingly, only GU and PNOF show a
pronounced step which compares well with the fundamental gap. The other
functionals show either a completely smooth behavior or, in the case of BBC3, a
small kink in the wrong direction. Therefore, we conclude that the complete
removal of the  SI terms is essential for obtaining the correct behavior of
$\mu(M)$. The size of the step of $\mu(M)$ for the PNOF is $0.30$ Ha and
compares well with experiment (see Table~\ref{tab_gap}). As a test, we also
tried a modified version of BBC3 where we removed the SI terms completely.
Consistent with the above conclusion, it also produces a step which is almost
identical to PNOF.  Additionally, this modified BBC3, like the GU functional,
gives an accurate measure of the correlation  energy at the equilibrium
distance, but fails completely at the dissociation limit.

\section{Conclusion} 
We have presented a formalism to calculate the fundamental gap within RDMFT for
both open- as well as closed-shell systems. Our numerical results show that
even for systems where the neutral system is closed-shell the results for the
chemical potential are closer to the exact step function if an open-shell
treatment is employed because adding charge of a specific spin to the system
makes it open-shell. The application to several open-shell systems gives a very
good agreement with experimental values in all cases. Also, the steps in the
chemical potentials are such that they resemble the spin dependence of the
ionization potential and the electron affinity of the real  system. Our
investigation of a possible basis set dependence reveals that it is necessary
to include very diffuse states in the basis set in case the system does not
bind extra charge. Whenever the system does bind extra charge the results are
independent of the inclusion of the diffuse state in the basis set. To estimate
the contribution of the occupation numbers and the natural orbitals to the
fundamental gap we compared the results for the LiH molecule using a full
variation and a variation of the occupation numbers only. We found that over
90\% of the fundamental gap are due to the occupation numbers. This finding was
confirmed for several other systems so far and we believe that it shows a
general feature of RDMFT calculations. Finally, we investigated the behavior of
several different functionals for the calculation of the fundamental gap. From
our results we conclude that the exclusion of the self-interaction for all
natural  orbitals is essential to obtain reasonable results. Functionals
without any removal of self-interaction simply yield a continuous chemical
potential. 

The present work is a contribution to the subject of calculating the
fundamental gap of materials within RDMFT. The hope is that this theory gives
results closer to experiment than DFT for this fundamental problem. It is our
belief that the theoretical development presented in this work will have a
significant impact in the application of RDMFT to periodic systems.

\begin{acknowledgments} 
We would like to thank A. Zacarias for valuable discussions on experimental and
different theoretical results. This work was supported in part by the Deutsche
Forschungsgemeinschaft within the program SPP 1145,  and by EU's Sixth
Framework Program through the Nanoquanta Network of Excellence
(NMP4-CT-2004-500198). 
\end{acknowledgments}

\end{document}